\documentstyle[12pt,epsf]{article}
\begin{document}
\thispagestyle{empty}
\begin{flushright}
CUPP-99/6\\
\texttt{hep-ph/9909453} \\
September 1999\\
\end{flushright}
\vskip 5pt

\begin{center}
{\large {\bf VACUUM OSCILLATION SOLUTIONS OF THE SOLAR NEUTRINO
PROBLEM : A STATUS REPORT}}

\vskip 10pt

{\sf Srubabati Goswami$^a$
\footnote{E-mail address: sruba@prl.ernet.in}},  
{\sf Debasish Majumdar$^b$
\footnote{E-mail address: debasish@tnpdec.saha.ernet.in}},  
and 
{\sf Amitava Raychaudhuri$^b$
\footnote{E-mail address: amitava@cubmb.ernet.in}}  

\vskip 8pt
$^a${\em Physical Research Laboratory, Navrangpura,\\ 
Ahmedabad 380 009, India }\\

$^b${\em Department of Physics, University of Calcutta,\\ 
92 Acharya Prafulla Chandra Road, Calcutta 700 009, India }

\vskip 8pt

{\bf ABSTRACT}

\end{center}


We re-examine vacuum oscillation solutions of the solar neutrino
problem taking (a) the results on total rates, electron energy
spectrum, and the seasonal variations from the 708 day
SuperKamiokande data and (b) those on total rates from the
Chlorine and Gallium experiments.  Best fit values for the mixing
angle and mass splitting are found for oscillations to sequential
and sterile neutrinos and the 90\% C.L.  allowed regions are
determined.

\newpage

\section{Introduction}

Within the past year much has taken place in the arena of neutrino
physics. The announcement by SuperKamiokande (SK) of the strong
evidence in support of a non-zero neutrino mass and oscillation
in their atmospheric $\nu$ data \cite {ska} has catalyzed
explorations of the subject from many different angles. The
far-reaching impact of a neutrino mass on physics beyond the
Standard Model and in astrophysics and cosmology  \cite {numass}
is being vigorously examined. Additional details of the neutrino
mass and the determination of the complete mass spectrum are
therefore awaited with  interest.

The results on the arrival rates of solar neutrinos \cite{solar}
from the Chlorine, Kamiokande and Gallium experiments were
themselves strong indications of a non-zero neutrino mass. Recent
high statistics results on solar neutrino rates, the scattered
electron spectrum, and the seasonal variation of the rates from
SuperKamiokande have added a new dimension to this effort
\cite{sksolar}.  In this work we consider the vacuum oscillation
scenario in the light of this body of data from several
directions \footnote {Prior to SK, the solar neutrino data have
been examined in terms of vacuum oscillation by many authors
\cite{preskvac}.}.  We utilize the latest 708 day SK data
\cite{smy} and for the
neutrino fluxes use the BP98 solar model \cite{BP98} which
incorporates the INT normalisation.  We consider the two
alternatives of oscillation of the $\nu_e$ to a sequential
($\nu_\mu$ or $\nu_\tau$) or a sterile neutrino.

\section{Oscillation Probability}

In this work we restrict ourselves to the simplest case of
mixing between two neutrino flavors.
The $\nu_e$ survival probability used by us is
\begin{equation}
P_{ee} (E_\nu,r,R(t)) = 1 -
\sin^{2}{2\vartheta}\sin^2\left[{\frac{\pi
R(t)}{\lambda}}\left(1-\frac{r}{R(t)}\right)\right]
\label{p2vacn}
\end{equation}  
where $\vartheta$ denotes the mixing angle in vacuum and
$\lambda$ is the vacuum oscillation wavelength for neutrinos of
energy $E_\nu$ given by $4 \pi E_\nu/\Delta m^2$, in which
$\Delta m^2 = |m_1^2 - m_2^2|$ is the mass square splitting. Here
$r$ is the distance of the point of neutrino production from the
center of the sun and $R(t)$ is the sun-earth distance given by,
\begin{equation}
R(t) = R_{0} \left[ 1 - \epsilon \cos(2 \pi \frac{t}{T})\right]
\end{equation}
Here, $R_{0} = 1.49 \times 10^{13}$ cm is the mean Sun-Earth
distance and $\epsilon = 0.0167$ is the ellipticity of the
earth's orbit. $t$ is the time of the year at which the solar
neutrino flux is measured and $T$ is 1 year.

\section{Observed rates and neutrino oscillations}

After the declaration of the SK results, analysis of the total
rates of all the experiments in terms of vacuum oscillation has
been considered in \cite{bks} using the 504 day data. In this
section we update this analysis by using the data collected over
708 days \cite{smy}.  The data that we use in our analysis for
the total rates are given in Table 1.  For the Ga experiments we
take the weighted average of the SAGE and Gallex results. Because
SK has better statistics we do not include the Kamiokande
results.  We fit the total rates from the three experiments using a
standard $\chi^2$-fitting procedure \cite{minuit}.

The definition of $\chi ^2$ used by us is,
\begin{equation}
\chi^2 =
\sum_{i,j=1,3} \left(F_i^{th} -
F_i^{exp}\right)
(\sigma_{ij}^{-2}) \left(F_j^{th} - F_j^{exp}\right)
\label{ratefit}
\end{equation}
Here $F_{i}^{\alpha}= \frac{T_i^{\alpha}}{T_{i}^{BP98}}$ where
$\alpha$ is $th$ (for the theoretical prediction) or $exp$ (for
the experimental value) and $T_i$ is the total rate in the $i$th
experiment.  $F_{i}^{exp}$ is taken from Table 1.  The error
matrix $\sigma_{ij}$ contains the experimental errors, the
theoretical errors and their correlations.  For evaluating the
error matrix we use the procedure described in \cite{flap}.  In
the presence of neutrino conversions, the detection rate on earth
for the radiochemical experiments $^{37}{Cl}$ and $^{71}{Ga}$ is
predicted to be:
\begin{equation}
(T)_{\i }^{th} =
\sum_k \int_{E_{th}} \phi _k(E_\nu) \sigma (E_\nu)
<P_{ee}(E_\nu,r,R(t))> dE_\nu
\label{eq:final}
\end{equation}
where $\sigma (E_\nu)$ is the cross-section for neutrino capture
\cite{jnbhome} and $\phi_{k}(E_\nu)$ is the neutrino spectrum for the
$k$th source \cite{jnbhome}.  The sum is over all the individual
neutrino sources.  $<P_{ee}(E_\nu,r,R(t))>$ is the average neutrino
survival probability over one year for the time averaged total
rates and further where a weighted sum over the production point
in the sun has been carried out for each source.  The theoretical
prediction according to the BP98 solar model, $T_{i}^{BP98}$, is
obtained by setting the survival probability  as 1 in the above.
For SK, in the case of oscillation to sequential neutrinos one
has to take into account the possible contributions from the
$\nu_\mu$ or $\nu_\tau$ channel,
\begin{eqnarray}
(T)_{i}^{th} & = &
\int_{E_{A_{th}}} dE_{A} \int_{0}^{\infty} dE_{T} 
\rho(E_{A}, E_{T})
\int_{E_{\nu_{min}}}^{E_{\nu_{max}}} dE_\nu \phi _i(E_\nu) \nonumber \\
&  & \left[<P_{ee}(E_\nu,r,R(t))> \frac{d\sigma_{\nu_e}}{dE_{T}} +
<P_{e\mu}(E_\nu,r,R(t))> \frac{d\sigma_{\nu_\mu}}{dE_{T}}\right]
\label{rkam}
\end{eqnarray}
The second term in the bracket is absent if oscillation to
sterile neutrinos is under consideration.  $E_{T}$ and $E_{A}$
denote the true and apparent electron energies respectively. $\rho
(E_{A}, E_{T})$ is the energy resolution function for which we
use the expression given in \cite{blk}.  $E_{A_{th}}$ is 6.5 MeV
for the calculation of the total rate at SK.
$\frac{d\sigma}{dE_{T}}$ is the differential cross-section for
the production of an electron with true relativistic energy
$E_{T}$ and can be calculated from standard electroweak theory.
Below we summarize our results for the total rates. For the
sequential neutrino case the best-fit values of $\Delta m^2$,
$\sin^2(2\vartheta)$  and $\chi^2_{min}$
obtained are
\begin{itemize}
\item{$\Delta m^2  = 9.5 \times 10^{-11}$ eV$^2,\;\;\;\;
\sin^2 (2\vartheta) = 0.87,\;\;\;\;
\chi^2_{min} = 0.52$}
\end{itemize}
For 1 degree of freedom (3 experimental data points -- 2 parameters)
this solution is allowed at 47.08\% C.L. 
The best-fit point that we find is slightly different from that in
\cite{bks}. We attribute this to the fact that in the above analysis we
have not included the detector efficiencies. Incorporating these
\cite{zconer}, we get 
\begin{itemize}
\item{$\Delta m^2  = 7.8 \times 10^{-11}$ eV$^2,\;\;\;\;
\sin^2 (2\vartheta) = 0.74,\;\;\;\;
\chi^2_{min} = 2.54$}
\end{itemize}
These values are in agreement with \cite{bks} and with the 708
day data the quality of the fit is actually better. In fig. 1 we
show the 90\% C.L.  ($\chi^2 \leq \chi^2_{min}$+ 4.61) contours
for the vacuum oscillation solution. Apart from the global
minimum there are several local minima as shown in the figure.
The best fit for the sterile neutrino alternative gives $\chi^2 =
5.74$, which, for 1 degree of freedom, is ruled out at 98.3\%
C.L.

\section{Observed spectrum and neutrino oscillations} 
In addition to the total rates, SK has provided the number of
events in 17 electron recoil energy bins of width 0.5 MeV in the
range 5.5 MeV to 14 MeV and an 18th bin which covers the events
in the range 14 to 20 MeV \cite{smy}.  In this section we analyze
the spectral data in the light of neutrino oscillations in
vacuum.  The definition of $\chi ^2$ used by us is,
\begin{equation}
\chi^2 =
\sum_{i,j=1,18} \left(X_{n} R_i^{th} -
R_i^{exp}\right)
\sigma_{ij}^{-2} \left(X_n R_j^{th} - R_j^{exp}\right)
\label{spectfit}
\end{equation}
where $X_n$ allows for  an arbitrary  normalisation of the $^{8}{B}$
flux with respect to the SSM prediction and $R_i^{\alpha} =
S^{\alpha}_i/S_i^{BP98}$ with $\alpha$ being $th$ or $exp$ as before
and $S_i$ standing for the number of neutrinos in the $i$th energy bin.
The theoretical prediction is given by eq. (\ref{rkam}) but the
integration over the apparent ({\em i.e.}, measured) energy will now be
over each bin.  
Following SK we include the statistical error, the uncorrelated
systematic errors and the energy-bin-correlated experimental errors 
as well as those from the calculation of the expected spectrum
\cite{skspec}.
Thus the error matrix $\sigma_{ij}$ used by us is
\begin{equation}
\sigma_{ij}^2 = \delta_{ij}(\sigma^2_{i,stat} +
\sigma^2_{i,uncorr}) + \sigma_{i ,exp} \sigma_{j,exp} +
\sigma_{i,cal} \sigma_{j,cal}
\end{equation}
Our results are given below:\\
(a) Oscillation to a sequential neutrino\\
The best-fit values of parameters and $\chi^2_{min}$ are
\begin{itemize}
\item
{$\Delta m^2  = 4.15 \times 10^{-10}$eV$^2,\;\;\;
\sin^2 (2\vartheta) = 0.89,\;\;\;
X_n =0.75,\;\;\;
\chi^2_{min}$ = 13.07}.
\end{itemize}
This solution is allowed at 59.68\% C.L. The allowed values of the 
parameters $\Delta m^2$ and $\sin^2 (2\vartheta)$ at 90\% C.L.
($\chi^2 \leq \chi^2_{min}$ + 6.25) are shown in fig. 2a.
The best-fit values of $\Delta m^2$ and $\sin^2 (2\vartheta)$
that we get are in agreement with those obtained in \cite{smy}
for vacuum oscillations.\\ (b) Oscillation to a sterile
neutrino\\ For this case we obtain,
\begin{itemize}
\item{$\Delta m^2  = 4.16 \times 10^{-10}$eV$^2,\;\;\;\;\;
\sin^2 (2\vartheta) = 0.77,\;\;\;
X_n = 0.76,\;\;\;\;\;\;
\chi^2_{min}$ = 12.9}
\end{itemize}
This corresponds to a  goodness of fit of 61\%.  Thus, as far as the
electron recoil spectrum data is considered the sterile neutrino
alternative gives a marginally better fit.  In fig. 2b we show the
allowed values of parameters for this case at 90\% C.L.\footnote {In the
analysis of the spectrum data we have not included the efficiencies as
they are available in \cite{zconer} as functions of true energy for
particular values of the threshold apparent energy. Here, we perform 
bin by bin integration over the apparent energy and there are different
thresholds for each bin for which the appropriate efficiencies are
unavailable. We do not have any reason to believe that the
inclusion of the efficiencies will change the best-fit values
dramatically -- see {\em e.g.} the fits in Sec. 3 -- though the
quality of the fit may be affected.}

\section{Combined fits to rates and spectrum}
In this section we present the results of the combined fit of the total
rates and the spectrum data treating the rates and the electron
spectrum data as independent \cite{bks}.  Thus our definition of $\chi
^2$ is,
\begin{eqnarray}
\chi^2 &=&
\sum_{i,j=1,3} \left(F_i^{th} -
F_i^{exp}\right)
\sigma_{ij}^{-2} \left(F_j^{th} - F_j^{exp}\right) \nonumber \\
&& + \sum_{i,j=1,18} \left(X_n R_i^{th} -
R_i^{exp}\right)
\sigma_{ij}^{-2} \left(X_n R_j^{th} - R_j^{exp}\right)
\end{eqnarray}
where the first term on the r.h.s is from eq. (\ref{ratefit}) and
the second from eq. (\ref{spectfit}).
We allow the normalisation of the $^{8}{B}$ flux to vary as a free
parameter.  The $\chi^2_{min}$ and the best-fit values we obtain
are:\\
(a) Sequential neutrino case
\begin{itemize}
\item{$\Delta m^2  = 9.47 \times 10^{-11}$ eV$^2,\;\;\;\;\;
\sin^2 (2\vartheta) = 0.80,\;\;\;
X_n = 0.71,\;\;\;\;
\chi^2_{min} = 18.66$}
\end{itemize}
For 18 (= 21 -- 3) degrees of freedom this solution is allowed at
41.3\% C.L.. In fig. 3a we show the 90\% C.L. allowed regions for the
combined analysis of rate and spectrum.\\ 
(b) Sterile neutrino case
\begin{itemize}
\item{$\Delta m^2  = 8.86 \times 10^{-11}$ eV$^2,\;\;\;\;\;
\sin^2 (2\vartheta) = 0.88,\;\;\;
X_n = 0.82,\;\;\;\;\;
\chi^2_{min} = 18.54$}
\end{itemize}
The goodness of fit in this case is 42.06\%. 
In fig. 3b we show the 90\% C.L. allowed region for oscillations
to a sterile neutrino from the combined analysis of rates and
spectrum.  We find that only two of the six regions allowed for
the sequential case remain for the sterile alternative while for
the others either the $\chi^2$ is too high or they merge with
these two.

\section{Seasonal variation and neutrino oscillations}
In \cite{smy} the SK collaboration has also presented the preliminary
seasonal data which shows a variation apart from that expected from 
the $1/R^2(t)$ dependence of the neutrino fluxes. 
If this is confirmed then it can help to distinguish between the 
MSW \cite{msw} and the vacuum oscillation alternatives. 
To analyze the seasonal data we define our $\chi^2$ as  
\begin{equation}
\chi^2 =
\sum_{i,j=1,8} \frac{\left(\frac{N_i^{th}}{N_{i}^{BP98}} -
\frac{N_i^{exp}}{N_{i}^{BP98}}\right)^2}{\sigma_{i}^2}
\end{equation}
$N_{i}^{th}$ is as given by eq. (\ref{rkam}) but now the
integration over the apparent energy is from 11.5 MeV. 
The time averaged probability for this case is 
\begin{equation}
\langle{P_{ee}(E_\nu,r,R(t)}\rangle = \frac{1}{t_2 - t_1}
\int_{t_1}^{t_2} {P_{ee}(E_\nu,r,R(t)) dt}
\end{equation}
where $t_2$ and $t_1$ are determined by the eight bins provided by
SK. For each of these bins, the SK collaboration has presented the
ratio $(N_i^{exp}/N_{i}^{BP98})$ \cite{smy}. The best-fit values for
oscillation to sequential neutrinos obtained in this case are
\begin{itemize}
\item{$\Delta m^2 = 4.26 \times 10^{-10}$ eV$^2,\;\;\;\;\;
\sin^2 (2\vartheta) = 1.0,\;\;\;
X_n = 1.12,\;\;\;\;
\chi^2_{min} = 3.96$ }
\end{itemize}
The goodness of fit is 55.52\%.  We find that from the seasonal data
almost all of the parameter space in the $\Delta m^2 - \sin^2
(2\vartheta)$ plane is allowed at 90\% C.L..  In fig. 4 we plot
$\chi^2_{min}$ against one of the three parameters keeping the other
two unconstrained to illustrate this point.  From fig. 4 it is seen
that  all values of $\Delta m^2$ in the range $10^{-11} - 10^{-9}$
eV$^2$ and $\sin^2 (2\vartheta)$ from very small values to 1.0 are
allowed at 90\% C.L..  $X_n$ below 0.5 and above 3.5 are not allowed at
90\% C.L..  Thus, the preliminary seasonal variation data does not put
any strong constraint on the parameters.  For oscillation to sterile
neutrinos the best-fit values are
\begin{itemize}
\item{$\Delta m^2 = 4.09 \times 10^{-10}$ eV$^2,\;\;\;\;
\sin^2 (2\vartheta) = 1.0,\;\;\;\;
X_n = 0.99,\;\;\;\;\;
\chi^2_{min} = 3.88$}
\end{itemize}
For 5 (8 -- 3) degrees of freedom this solution is allowed at 56.68\%. 

\section{Conclusions}

In this paper a detailed $\chi^2$-analysis of the 708 day SK
solar neutrino data is performed assuming vacuum oscillation
between two neutrino flavors.  This includes
\begin{itemize}
\item{A $\chi^2$-analysis using the total rates from the $^{37}{Cl}$,
$^{71}{Ga}$ and Superkamiokande experiments. We take into account the
experimental and the theoretical errors including the correlations
among the various theory errors.}
\item{A $\chi^2$-analysis using the SK spectrum data including the
uncorrelated as well as the correlated errors among various bins as
given by the SK collaboration \cite{sksolar}.  We float the
normalisation of the $^{8}{B}$ flux as a free-parameter and determine
its best-fit value.}
\item{A global $\chi^2$-analysis of the combined rates and spectrum
data treating them as independent. For this case also we have allowed
the $^{8}{B}$ flux to vary as a free-parameter.}
\item{$\chi^2$-analysis of the preliminary seasonal data allowing the
$^{8}{B}$ flux normalisation to vary. Here we consider only
the experimental statistical errors.}
\end{itemize}  

We find that the simple two-generation vacuum oscillation
scenario can well explain the data on total rates and the
spectrum but somewhat different best-fit values of $\Delta m^2$
are found in the two cases.  If one does a global analysis of the
rates and the spectrum then the goodness of fit is 
poorer. This is because of the fact that the best-fit value of
$\Delta m^2$ for the spectrum data is one order of magnitude
higher as compared to the rates value.  The sterile neutrino
alternative gives a bad fit to the data on total rates but for
the SK spectrum data as well as for the combined rate and
spectrum data it gives a marginally better fit than the active
neutrino scenario. The sterile neutrino also gives a
slightly better fit for the preliminary seasonal variation data.
Thus we make an important observation that {\it if only the SK
spectrum or the seasonal data are considered then the sterile
neutrino actually gives a slightly better fit than the sequential
neutrino case.} It is when the rates in the radiochemical
experiments are included that the sterile neutrino fit becomes
worse.

Two-generation vacuum oscillation analysis of the 708 days of 
SK data has also been performed in
\cite{bw2}. However their fitting procedure is somewhat different from
ours.  We take into account the total SK rate as well as the spectrum
data.  We also include the theory errors and their correlations in the
analysis of the total rates.  For the SK spectrum data we have taken
the bin by bin correlations into account.  Since they have not included
the SK rates we cannot compare our results for the total rates with
theirs.  If we compare our results of only the spectral and seasonal
data with their analysis, the best-fit values are more or less in
agreement though our $\chi^2_{min}$ are somewhat lower. Because the
seasonal data does not put strong constraints on the parameters,
this preliminary data have not been included in our global fit.
We have not included the data from the measurement of day-night 
solar neutrino flux \cite{dn} as the vacuum oscillation hypothesis 
does not generate any day-night asymmetry. 

In order to explain the energy spectrum observed at SK, in the
literature the $hep$ flux has sometimes been allowed to vary
apart from the $^{8}{B}$ flux \cite{bk} and it was found that a
$hep$ flux almost 20-30 times larger than the SSM prediction can
well explain the high energy part of the spectrum data. In this
work we have not considered this possibility.  Whether one
requires such a higher $hep$ flux than the SSM prediction can be
tested in SK. It was concluded in \cite{bw2} that even if one
allows the $hep$ flux to vary, the allowed parameters for the
vacuum oscillation hypothesis will not change much.
 
For our global analysis of the rates and the spectrum we have treated
these two data sets as independent. Since the $^{8}{B}$ neutrino flux
enters the rates as well as the spectrum data there can be some
possible correlations among these.  In a future study  we plan to
examine whether the inclusion of these alters the conclusions 
by a significant amount.

\parindent 0pt

{\large{\bf {Acknowledgements}}}

D.M. and A.R. are partially supported by the Eastern Centre for
Research in Astrophysics, India. A.R. also acknowledges a
research grant from the Council of Scientific and Industrial
Research, India. S.G. is grateful to Saha Institute of
Nuclear Physics for hospitality for a period when the work was in
progress. She would also like to thank Plamen Krastev for many 
useful correspondences.

\newpage

Table 1: The ratio of the observed solar neutrino rates to the
corresponding BP98 SSM predictions used in this analysis. The results
are from Refs. \cite {solar} and \cite {smy}.  For Gallium, the
weighted average of the SAGE and Gallex results has been used.

\begin{center}
\begin{tabular}{|c|c|c|c|}
\hline
Experiment & Chlorine & Gallium & SuperKamiokande \\ \hline
$\frac{\rm Observed \;\; Rate}{\rm BP98 \;\; Prediction}$ & 
$0.33 \pm 0.029$ & $0.57 \pm 0.054$ & $0.471 \pm 0.015$ \\ \hline
\end{tabular}
\end{center}
~
\begin{center}
{\bf{\Large Figure Captions}}
\end{center}

Fig 1. The 90\% C.L. allowed region in the $\Delta m^2$ - $\sin^2
(2\vartheta)$ plane from the analysis of total rates assuming vacuum
oscillations to sequential neutrinos. The best-fit point is also
indicated. \\ \\
Fig 2. The 90\% C.L. allowed region in the $\Delta m^2$ - $\sin^2
(2\vartheta)$ plane from the SK recoil electron spectrum data for
sequential neutrinos (a) and for sterile neutrinos (b). The
best-fit points are also indicated. \\ \\
Fig. 3. The 90\% C.L. allowed region in the $\Delta m^2$ - $\sin^2
(2\vartheta)$ plane from a global analysis of the rates and the
spectrum data for sequential neutrinos (a) and for sterile neutrinos
(b).  The best-fit points are also indicated.\\ \\
Fig. 4. Variation of $\chi^2_{min}$ (solid line) (a) with $\Delta
m^2$ keeping $\sin^2 (2\vartheta)$ and $X_n$ unconstrained; (b)
with $\sin^2 (2\vartheta)$ keeping $\Delta m^2$ and $X_n$
unconstrained; and (c) with $X_n$ keeping $\sin^2 2\vartheta$ and
$\Delta m^2$ unconstrained.  Also shown are the 90\% C.L
(big-dashed line), 95\% C.L. (small-dashed line) and 99\% C.L.
(dotted-line) limits.

\begin{figure}[p]
\epsfxsize 15 cm
\epsfysize 15 cm
\epsfbox[25 151 585 704]{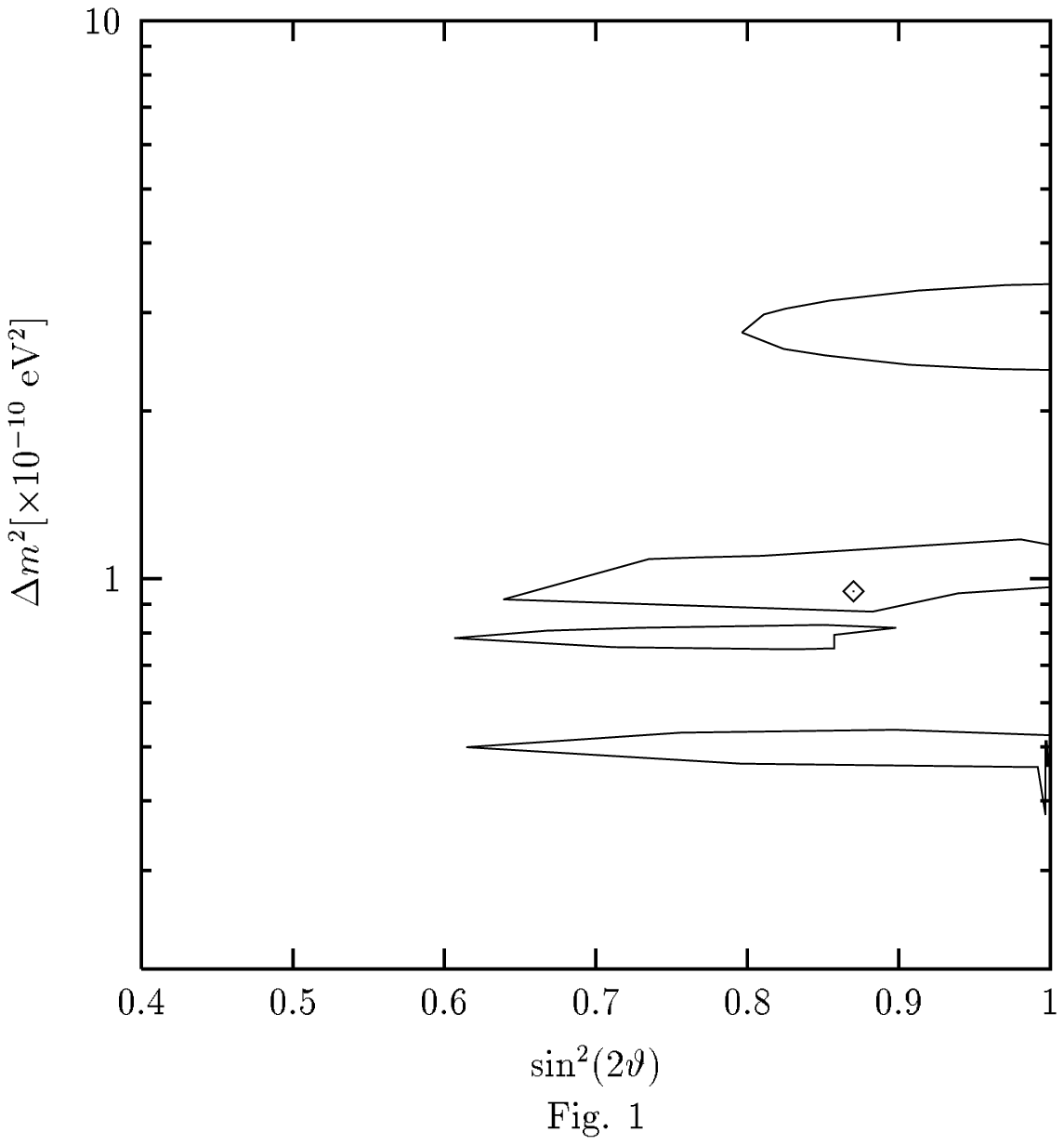}
\end{figure}

\begin{figure}[p]
\epsfxsize 15 cm
\epsfysize 15 cm
\epsfbox[25 151 585 704]{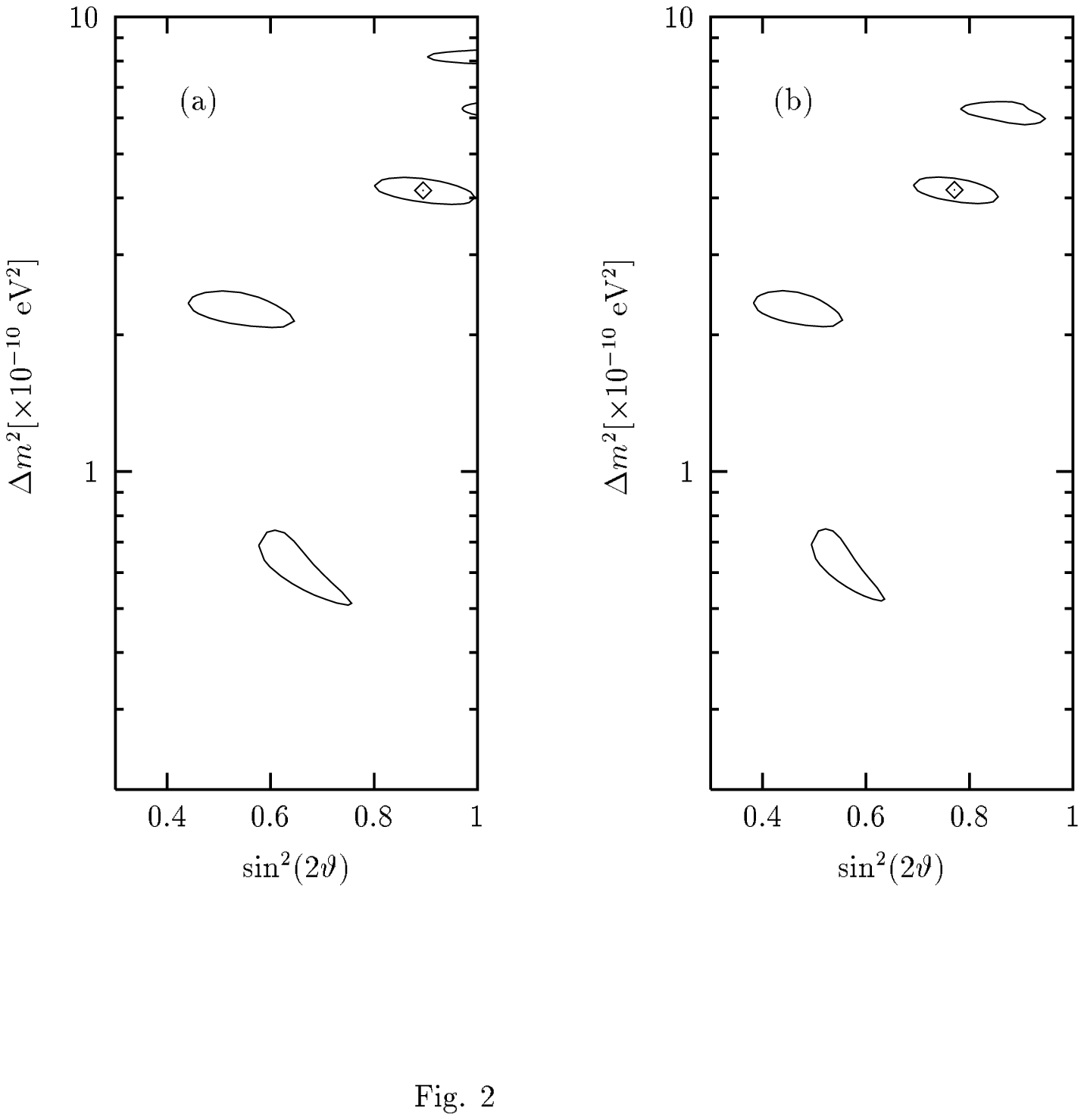}
\end{figure}

\begin{figure}[p]
\epsfxsize 15 cm
\epsfysize 15 cm
\epsfbox[25 151 585 704]{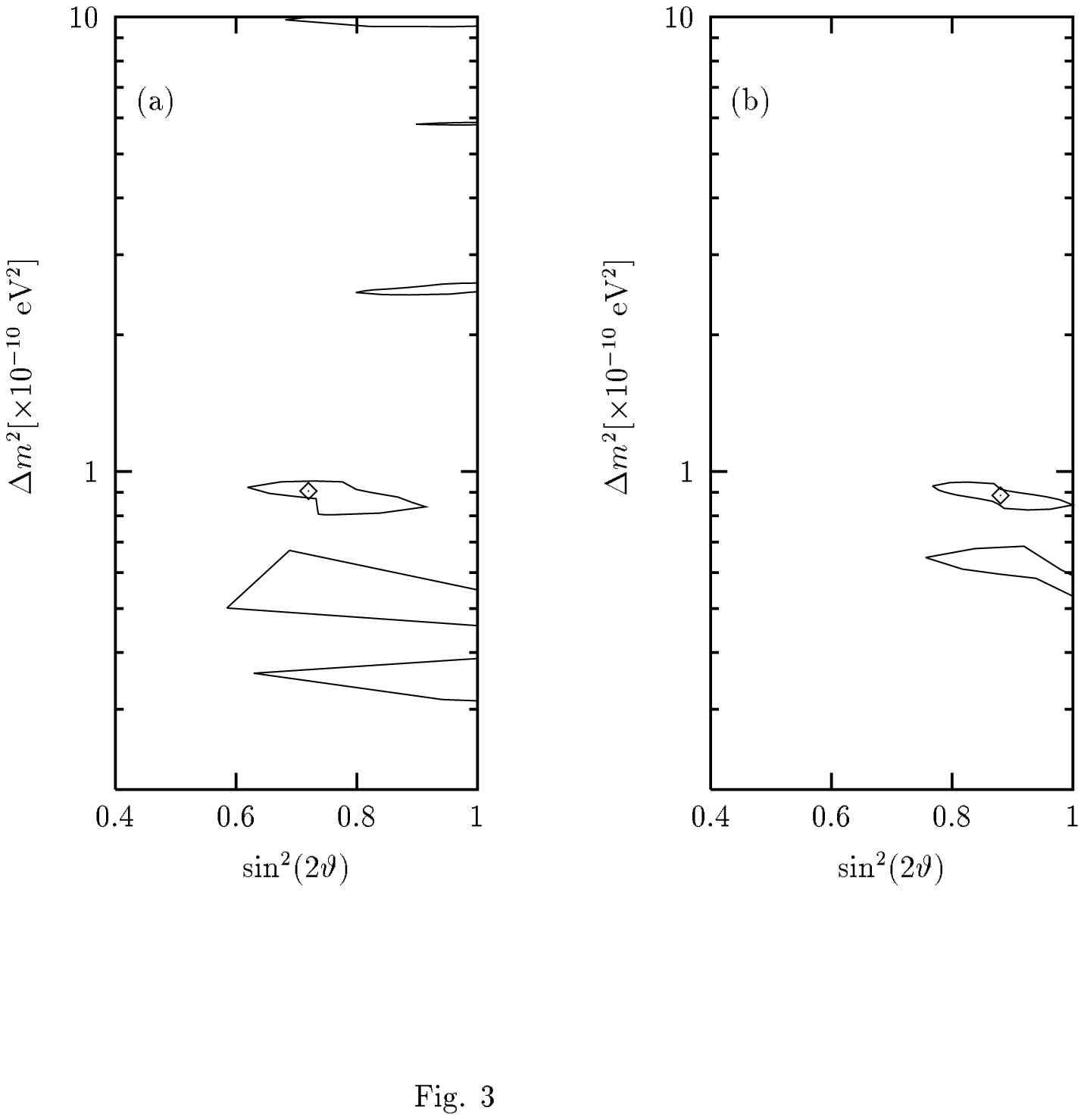}
\end{figure}

\begin{figure}[p]
\epsfxsize 15 cm
\epsfysize 15 cm
\epsfbox[25 151 585 704]{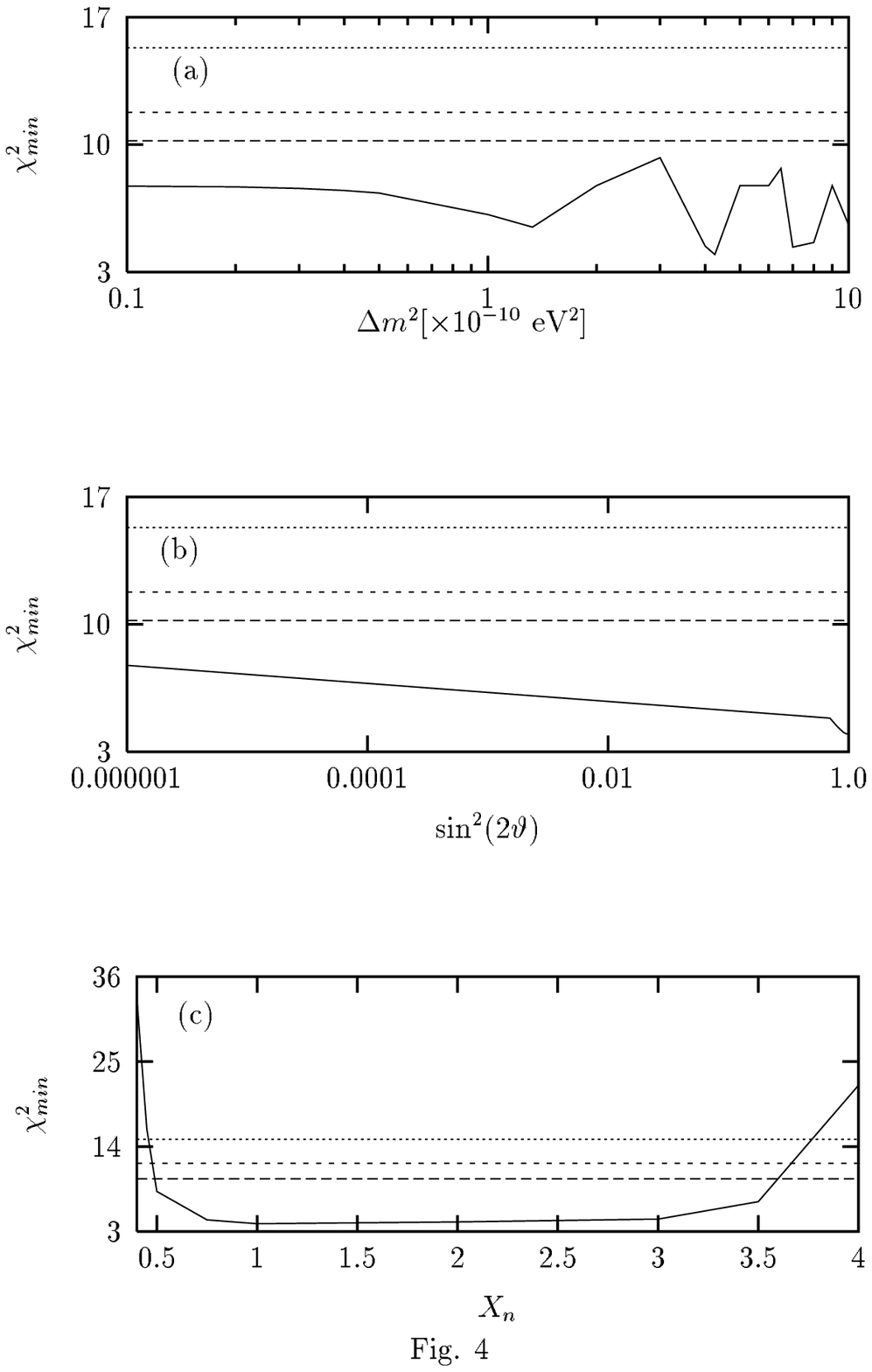}
\end{figure}

\end{document}